# Electronic Metamaterials with Tunable Second-order Optical Nonlinearities

Hung-Hsi Lin,[1†] Felipe Vallini,[2†] Mu-Han Yang,[2] Rajat Sharma,[2]
Matthew W. Puckett,[2] Sergio Montoya,[2,3] Christian D. Wurm,[2]
Eric E. Fullerton,[2,3] and Yeshaiahu Fainman[2*]

The ability to engineer metamaterials with tunable nonlinear optical properties is crucial for nonlinear optics. Traditionally, metals have been employed to enhance nonlinear optical interactions through field localization. Here, inspired by the electronic properties of materials, we introduce and demonstrate experimentally an asymmetric metal-semiconductor-metal (MSM) metamaterial that exhibits a large and electronically tunable effective second-order optical susceptibility ($\chi^{(2)}$). The induced $\chi^{(2)}$ originates from the interaction between the third-order optical susceptibility of the semiconductor ($\chi^{(3)}$) with the engineered internal electric field resulting from the two metals with dissimilar work function at its interfaces. We demonstrate a five times larger second-harmonic intensity from the MSM metamaterial, compared to contributions from its constituents with electrically tunable nonlinear coefficient ranging from 2.8 to 15.6 pm/V.

**Introduction**

Materials with large second-order nonlinear optical susceptibility ($\chi^{(2)}$) (e.g., crystals of lithium niobate ($LiNbO_3$) and potassium dihydrogen phosphate (KDP)) have been widely used for the realization of nonlinear optical functionalities including light modulation, switching and wave mixing.[1-4] However, integration of these materials with CMOS compatible photonic integrated circuits (PIC) remains challenging.[5,6] CMOS compatible materials are either amorphous (e.g., $SiO_2$, $HfO_2$ and $ZrO_2$, etc.)[7,8] or crystalline with a centrosymmetric diamond lattice structure (e.g., Si) and, consequently, do not possess second-order nonlinear optical susceptibility, $\chi^{(2)}$.[9] Strain engineering of crystalline silicon waveguides has been proposed and demonstrated as a promising mean to create a non-zero effective $\chi^{(2)}$, however, the magnitude of the strain induced $\chi^{(2)}$, has recently been shown to be much lower than that reported in the past (i.e., 8 ± 3 pm/V).[9-13] The electric field induced second-harmonic (EFISH) generation introduces an effective $\chi^{(2)}$ by the interaction of an externally applied electric field with the material's bulk third-order nonlinear susceptibility, $\chi^{(3)}$.[14] Nevertheless due to shielding of the external field by a semiconductor, a sufficiently high internal electric field exists only within the first tens of nanometers from the semiconductor interfaces for both metal-oxide-semiconductor (MOS) and metal-semiconductor (MS) structures.[15-17] Current research in creating CMOS compatible materials exhibiting a high $\chi^{(2)}$ coefficient mostly focuses on back-end processing[18], composite metamaterials that utilize alternative physical mechanisms to synthesize non-zero $\chi^{(2)}$,[19,20] metallo-dielectric multilayers,[21] dielectric-semiconductor multilayers,[22] all-dielectric composite metamaterials,[8,23] and nonlinear processes enhanced by plasmonic metasurfaces and nanoatennas.[24,25] However these approaches do not allow for a good control over the exhibited nonlinearities.

---

[1]Materials Science and Engineering, University of California, San Diego, 9500 Gilman Drive, La Jolla, California 92093, USA

[2]Department of Electrical & Computer Engineering, University of California, San Diego, 9500 Gilman Drive, La Jolla, CA 92093, USA

[3]Center for Memory and Recording Research, University of California, San Diego, 9500 Gilman Drive, La Jolla, CA 92093-0401, USA

[†]These authors contributed equally to this work

[*] email: fainman@eng.ucsd.edu

In this manuscript, we propose, systematically analyze and demonstrate experimentally an asymmetrical metal-semiconductor-metal (MSM) electronic metamaterial that exhibits a prominent and tunable effective $\chi^{(2)}$. The large induced $\chi^{(2)}$ originates from the interaction between the bulk third-order nonlinear optical susceptibility of the semiconductor, $\chi^{(3)}$ with the engineered non-zero net internal electric field created by the electronic transport close to the semiconductor interfaces clad with metals possessing dissimilar work functions. The MSM structure, consisting of CMOS-compatible amorphous silicon (a-Si) layer clad with aluminum (Al) and nickel (Ni), is shown experimentally to generate a second-harmonic field intensity five times larger than that from its constituents alone. Additionally, the nonlinear response of MSM is shown to be both proportional to the difference in the work functions of the cladding metals and actively controllable by an external electric field. The resulting effective $\chi^{(2)}_{zzz}$ tensor component ranges in value from 2.8 to 15.6 pm/V, making these structures suitable for on-chip optical switching and modulation.

**Analysis of the electronic nonlinear metamaterial**

The basic principle of the proposed MSM nonlinear optical metamaterial and the SEM image of the fabricated structure are shown in Fig. 1a, where a thin layer of a-Si is clad with Al and Ni which have respectively smaller and larger work function (i.e., Al with 4.08 eV and Ni with 5.01 eV),[26,27] compared to the electronic affinity of a-Si (i.e., 4.67 eV). Consequently, a static electric field is generated within the bulk of the a-Si semiconductor layer. The band diagram of the MSM structure is included as as an insert in Fig. 1b showing a-Si's energy band gap $E_g$, Fermi level $E_F$, valence and conduction bands energies $E_V$ and $E_C$ respectively. $\phi_{Al}$ and $\phi_{Ni}$, are the work function of Al and Ni respectively. As metals and a-Si are brought into contact, Al which has a lower work function causes diffusion of electrons into a-Si until thermal equilibrium is reached and the $E_F$ of a-Si and Al align. The charge redistribution results in a built-in or diffusive potential, and creates a built-in static electric field close to the Al-a-Si interface, bending the bands of a-Si downwards. On the other side, once Ni with a larger work function is in contact with the a-Si layer, the heterostructure becomes asymmetric with the bands bent upwards close to the Ni-a-Si interface, resulting in an asymmetric built-in potential within the a-Si layer, and consequently a non-zero internal electric field, $E_{DC}$. The difference in the metals' work functions and the Fermi level of a-Si determines the depletion width and hence the magnitude and penetration depth of the induced electric field which can range from tens to hundreds of nanometers. With an appropriate choice of the two metals and a semiconductor layer thickness as narrow as the created depletion region, a high static electric field can be induced in the entire a-Si layer, leading to a high effective second-order nonlinear optical coefficient, $\chi^{(2)}$, via the EFISH effect.

We use TCAD Silvaco simulation tool[28] to study the magnitudes and spatial distributions of the built-in electric fields inside two types of MSM metamaterial structures consisting of a 25 nm thick a-Si layer clad with 5 nm thick metals either symmetrically (i.e., Al) or asymmetrically (i.e., Al and Ni) (see Fig. 1b). As expected, the net electric field in the case of the symmetric MSM metamaterial is zero (blue line), whereas in the case of the asymmetric MSM stack, a net DC electric field of $3.5 \times 10^5$ V/cm (red line) can be achieved. Since the a-Si layer exhibits a large third-order nonlinear susceptibility, $\chi^{(3)}$ (~ $2.3 \times 10^{-19}$ to $9.2 \times 10^{-19}$ m$^2$/V$^2$)[29-31] the induced DC electric field is expected to result in a high effective $\chi^{(2)} = 3\chi^{(3)} E_{DC}$, ranging from 10 to 40 pm/V.

**Characterization of nonlinear coefficient in various MSM metamaterial compositions**

For an experimental validation of our hypothesis, we designed and fabricated the MSM metamaterial on fused silica substrates using magnetron sputter at room temperature (see Methods). We then experimentally estimate the induced $\chi^{(2)}$ coefficients in a-Si layer by measuring the generated SHG intensity and subsequently using the revised Maker fringes analysis described in details in the Methods.[32,33]

First, to ensure that the detected signal is generated via the second-order nonlinear process from the bulk of deposited a-Si semiconductor, we analyze the measured signals as a function of the input pump power for both the MSM metamaterial as well as single layers of the cladding metals Al and Ni on fused silica substrates (see Fig. 2a). The measured intensities from all samples are confirmed, as expected, to be quadratically dependent on the pump power, illustrating the fact that the detected signals indeed originate from the second-order nonlinear responses. The a-Si film and the substrate are also confirmed to provide zero SHG signal as a result of their null $\chi^{(2)}$. We then apply Marker fringes method to characterize the polarization dependence of the SHG for extracting the $\chi^{(2)}$ tensor coefficients. The measured p- and s- polarized second-harmonic responses as a function of the polarization angle of the input pump are shown, for all the three cases, in Fig. 2b and 2c, respectively. For both polarizations, the SHG intensities from the MSM are clearly much larger than that generated from a combination of Al and Ni layers on fused silica substrates, implying that, as expected, the second-order nonlinear response is dominated by the a-Si layer due to the existence of the non-zero built-in electric field.

To further confirm our theory, we fabricated one more MSM metamaterial sample, replacing the top clad of Ni with platinum (Pt). Pt, just like Ni, has a larger work function (i.e., 5.65 e.V.)[26,27] than that of a-Si, and hence is expected to result in a non-zero built-in electric field in the a-Si layer in combination with the bottom clad of Al. The intensities of measured SHG from the Pt / a-Si / Al composition are measured and compared with the previous samples (i.e, Ni / a-Si / Al), as shown in Fig. 3, under the same experimental conditions. The SHG signal from the bottom Al cladding layer (red boxes) is fixed in all cases, and the signals from the top metal cladding films are measured individually and represented by blue boxes. The differences between the measured signals from the MSM metamaterials and their metallic constituents correspond to the contribution from a-Si layer (yellow boxes). It is apparent from Fig. 3 that the SHG signal in both MSM samples are significantly larger than the contributions from individual responses of their respective constituents. The generated second-harmonic field intensity can in fact be five times larger than the SHG signals from the respective constituents (e.g., individual Al, Ni and Pt metal films). Furthermore, and confirming our hypothesis, the contribution from the a-Si layer is found to be proportional to the difference of work functions between bottom and top metals, i.e., proportional to the built-in electric field. It can be envisioned that similar to the work demonstrated in this study, MSM metamaterials can, in the future, be custom-synthesized by choosing an appropriate combination of clad metals and semiconductor layer.

Additionally, the components of $\chi^{(2)}$ tensor in these MSM metamaterials are also estimated. We consider the thin MSM structures as a uniform isotropic effective medium[34], and then calculate all the induced $\chi^{(2)}$ tensor components using the measured p- and s-polarized SHG signals for each of the MSM metamaterial samples (see Methods). The results are summarized in the Table I. Comparing the magnitude of all non-zero tensor components, it is apparent that the all-normal tensor component, $\chi^{(2)}_{zzz}$, which is parallel to the direction of the built-in electric field, is dominating, proving conclusively that the additional contribution to the second-order nonlinear response of an asymmetrical MSM metamaterial is truly a result of the engineered electronic band structure in the bulk of the a-Si layer.

Since our experiments were carried out using optical fields with photon energies larger than bandgap of a-Si, we need to consider the effect of generated free-carriers that may affect the magnitude of the induced effective $\chi^{(2)}$. The pump generated photocarriers could lower the Schottky barriers between the semiconductor and metals, and cause a reduction in the induced built-in electric field. To estimate the strength of this effect, we measured and analyzed the I-V behaviors of the MSM stack. The experiments show that the built-in barrier height within the a-Si layer, under illumination with an average power of 100 mW (i.e., the power used in our experiments on SHG), will be reduced by 15% compared to its value under no illumination. This reduction, resulting from the image-force effect as a consequence of the generated photocurrent, however results ina reduction of the built in electric field by no more than 10% for the average pump-powers employed in this study. Therefore, we conclude that the pump-generated photocarriers have a relatively minor effect on the measured effective $\chi^{(2)}$. It should be noted that the reported measured values of effective $\chi^{(2)}$ may increase by avoiding photocarrier generation using pump photon energies smaller than the bandgap of a-Si (e.g., 1550 nm range).

**Electrical control of MSM nonlinear response**

In addition to their passive optical behavior, MSM metamaterials also enable the active control of their second-order nonlinear optical response by applying an external electric field using the constituent metals as electrodes. To illustrate this, experiments were carried out on a MSM metamaterial sample fabricated with a thicker a-Si layer of 50nm, since the built-in electric field in the 25 nm thick a-Si layer is already close to the breakdown limit. The schematic of this active MSM structure is shown in Fig. 4a. An external DC voltage is applied across the a-Si layer, and the generated SHG signal is measured for bias voltages ranging from -2.5 V to +1.5 V. Fig. 4b, shows the measured SHG intensity varying quadratically as a function of the applied field[16], which is expected for a linear change in the second-order nonlinear susceptibility coefficient as a function of the applied bias. A positive voltage increases the generated SHG signal by inducing more asymmetry in band structure of the a-Si layer, until the built-in electric field reaches the breakdown limit, which in our case occurs at +1.5 V. For negative voltages, we observe a reduction in the SHG signal as a result of the reduction in the built-in electric field. At the flat-band voltage, which is negative in this case, the electric field inside the semiconductor is completely eliminated, removing any of its contribution to the observed second-harmonic response. It should be noted, that even at this voltage we observed a finite, non-zero, SHG signal, due to the two metal films. However, this is not the voltage corresponding to the absolute minimum observed second-harmonic signal. This, in fact, is achieved at a voltage of -0.75 V, where the generated second-harmonic field intensities from a-Si and the metals are out of phase and completely cancel each other. The measured intensity, however, still does not reach a zero value, which is attributed to the fact that the non-diagonal components of $\chi^{(2)}$ tensor still remain non-zero and result in a finite SHG signal. As the magnitude of the voltage applied is increased further, the SHG signal increases again until the breakdown field is reached, which occurs at -2.5 V. It is noteworthy that the induced effective dipoles in a MSM metamaterial under negative and positive fields should exhibit opposite directions even though the measured SHG signal shows the same magnitude. The minimum and maximum values of the measured SHG intensity are then used to determine that a wide tunable range of 2.8 to 15.6 pm/V is achievable for of the effective $\chi^{(2)}_{zzz}$ tensor in this MSM metamaterial.

**Conclusions:**

Asymmetrical metal-semiconductor-metal (MSM) electronic metamaterials exhibiting a prominent and tunable effective second-order nonlinear optical susceptibilities $\chi^{(2)}$ are introduced and experimentally validated. It is conclusively shown that a large induced $\chi^{(2)}$ originates from the interaction between the bulk third-order nonlinear optical susceptibility of the semiconductor layer, $\chi^{(3)}$ with the engineered non-zero net internal electric field created by the electron transport on its interfaces with metals possessing dissimilar work functions. The MSM structures, consisting of CMOS compatible a-Si layer clad with Al-Ni and Al-Pt, are shown to generate as much as five times larger second-harmonic field intensity compared to the contribution from its constituents. The nonlinear response of MSMs, proportional to the difference in the work functions of the cladding metals, is shown to be actively controlled by external electric field resulting in tunable effective $\chi^{(2)}_{zzz}$ tensor component ranging from 2.8 to 15.6 pm/V, making it suitable for on-chip optical switching and modulation. Further enhancement in the effective $\chi^{(2)}$ of the MSM metamaterials can be achieved by using a semiconductor possessing a larger $\chi^{(3)}$, with a larger breakdown voltage, and/or by improving the quality of a-Si fabrication process to increase its breakdown voltage. Moreover, by varying the externally applied electric field, a tunable nonlinear effect is demonstrated, making the MSM metamaterial suitable for all-optical spatial signal processing.

**Methods**

**Sample preparation**

The metal/a-Si/metal metamaterials are grown at room temperature in an ultra-high vacuum environment by magnetron sputtering. Samples are deposited from Si-undoped (99.999%), Pt (99.95%), Ni (99.95%) and Al (99.95%) pure elemental targets onto fused silica substrates under an Ar pressure of 3 mTorr. Substrates undergo a thorough cleaning processes with Acetone, Isopropyl alcohol and deionized water before deposition, to prevent any inaccuracy in the optical characterization due to adhesion of particles on the surfaces. The thickness and refractive index of films are measured with the Rudolph Auto EL Ellipsometer. The deposition rate of each target and crystallization of films are characterized by means of small angle x-ray reflectivity (XRR) with a Bruker D8 Discover x-ray diffractometer. The XRR measurements, shown in the supplementary material of our previous work, confirm the fact that all sputtered films are amorphous.[22]

**Experimental methods**

The optical characterization is carried out via the Maker fringe setup that is used in our previous works.[22] The pump beam is generated using a Ti:Sapphire laser emitting 150 fs pulses with a 80 MHz repetition rate at a center wavelength of 800 nm. The polarization state of the pump is defined by a half-wave plate and a long pass filter with cut-off wavelength at 780 nm is set to filter out any signals from other sources in the range of interest. The sample is tilted and fixed at an angle of 45 degrees normal to the incident beam, which is focused onto the sample surface using an 10x objective lens, resulting in a beam size with a radius of 20 μm. At the output, two short wavelength pass filters and one band pass filter with a total optical density of 12 are inserted to filter out the pump light at ω (i.e., 800nm wavelength), ensuring that all photons collected by the photomultiplier (PMT) are at 2ω (i.e., 400 nm wavelength) and consequently generated from the SHG process. The collimated SHG signal from the sample is separated into p-polarized and s-polarized by a polarizer for calculating the different components of the $\chi^{(2)}$ tensor. The detected signal in the PMT is then read with an oscilloscope. A commercial 500 μm thick X-cut quartz wafer, exhibiting a nonlinear coefficient $\chi^{(2)}_{xxx}$ of 0.64 ± 8% pm/V is used to calibrate the system, and the absolute values of $\chi^{(2)}$ tensor components from our samples are determined by comparing the generated SHG signals with those from the quartz sample under the same experimental conditions. Bare fused silica substrates are also characterized under the same conditions as samples with grown metamaterials to ensure that substrates do not contribute to any SHG signal. The second-harmonic response from a single layer of a-Si is found to be negligible compared to the large detected signals from the metal films and MSM metamaterials which is expected due to its amorphous nature. The measurement errors in our setup originate mainly from the fluctuation of laser power due to the varying humidity in the environment (±5%), background noises (±20%), and the non-uniformity in the thickness of the deposited thin films (±10%). In addition, the possibility of counting error (±10%) of photons in the PMT (Hamamatsu Inc., H11461-03) due to pulse-overlapping, as described in the handbook, is also taken into account. In order to minimize these errors, the generated SHG intensities from quartz, metal films and MSM metamaterials are determined by taking the average of those measured from five different spots on each sample. Following Herman's equation,[32] three tensor components, $\chi^{(2)}_{zzz}$, $\chi^{(2)}_{xxz}$ and $\chi^{(2)}_{zxx}$ can be extracted by fitting the generated s- and p-polarized second-harmonic signals under various polarization angles of the incident pump beam.

**Maker fringes analysis**

Since a-Si is an amorphous material, we assume that its third-order susceptibility tensor $\chi^{(3)}$ components are same as those of an isotropic material with $C_{\infty,v}$ space symmetry and thus has 21 nonzero elements, of which only 3 are independent:

$$yyzz = zzyy = zzxx = xxzz = xxyy = yyxx \tag{1}$$

$$yzyz = zyzy = zxzx = xzxz = xyxy = yxyx \tag{2}$$

$$yzzy = zyyz = zxxz = xzzx = xyyx = yxxy \tag{3}$$

$$xxxx = yyyy = zzzz = zzxx + xzxz + xzzx \tag{4}$$

The existence of a static electric field in z direction is expected to introduce effective $\chi^{(2)}$ tensor components: $\chi^{(2)}_{xxz}$, $\chi^{(2)}_{xzx}$, and $\chi^{(2)}_{zxx}$ and $\chi^{(2)}_{zzz}$ through the EFISH effect. We assume that the MSM structures are isotropic in the transverse (i.e., in-plane) direction and that multiple reflections within the thin films can be neglected due to the thin nature of constituent films (i.e., 5 and 25 nm) compared to the wavelength of pump light (i.e., 800nm). With the assistance of Maker fringes analysis,[32,33] these non-zero $\chi^{(2)}$ tensor components of the MSM metamaterial: $\chi^{(2)}_{xxz}$, $\chi^{(2)}_{zxx}$ and $\chi^{(2)}_{zzz}$ can be determined from fitting the generated s- and p- polarized second-harmonic signal intensities at frequency 2ω as a function of the polarization of the fundamental pump beam at frequency ω, measured at a fixed angle of incidence, θ. Also, since the thicknesses of MSM metamaterials are much smaller than the wavelength of the pump beam, it is justified to use the effective medium theory for determining the three non-zero components of $\chi^{(2)}$ tensor[34] in all MSM metamaterial structures. The details in calculations are shown in the supplementary of our previous work.[22]

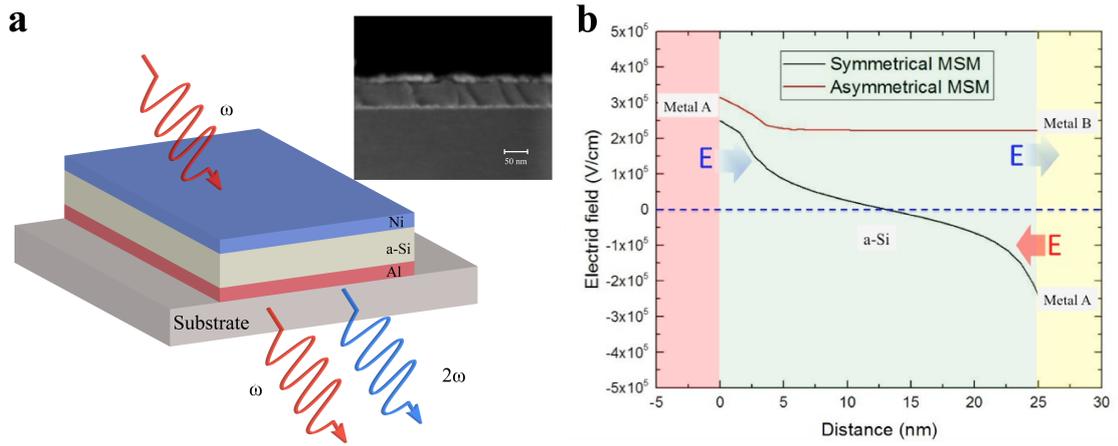

Fig. 1 (a) Schematic and SEM image of a 5 nm Ni / 25 nm a-Si / 5 nm Al MSM metamaterial for second-harmonic generation process. (b) The distribution of built-in electric field in symmetrical (black solid line), asymmetrical MSM structures (red solid line), and net electric field in symmetrical MSM (blue dash line) simulated with Silvaco.

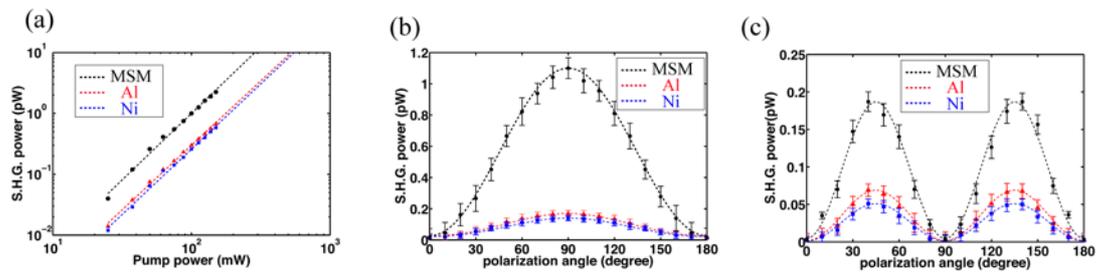

Fig. 2 (a) The log-log plot of measured SHG signal versus pump power for a single layer of Al (red triangles), Ni (blue squares) film and MSM metamaterial (black circles) grown on silica substrates for a fixed pump-beam incident angle of 45°. (b) The generated p-polarized and (c) s-polarized SHG intensity versus polarization angle for a 100 mW pump.

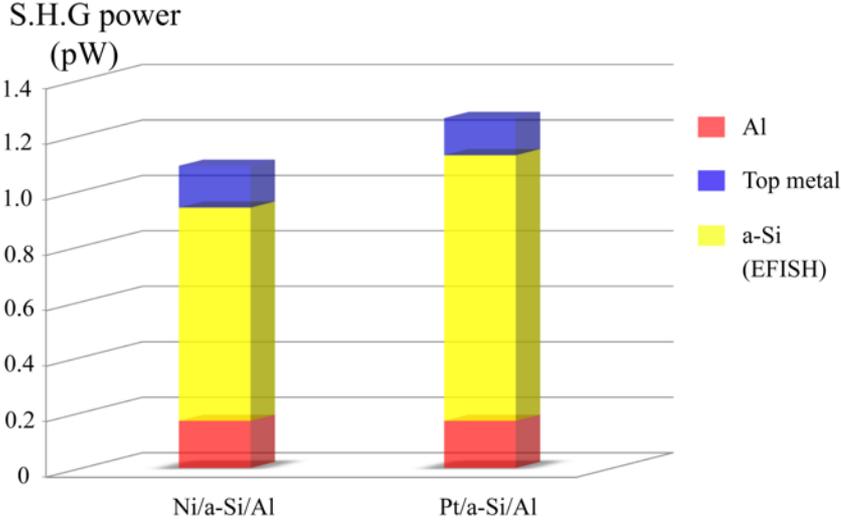

Fig. 3 Measured SHG intensities from the Ni/a-Si/Al (left) and Pt/a-Si/Al (right) MSM metamaterials and their constituents. The red boxes show the SHG from a single layer of Al film, and the blue boxes show the SHG from top cladding metal films. The difference between the total measured SHG intensities from the MSM metamaterials and their components is then assumed to be from the bulk of the a-Si (yellow) layer via EFISH effect.

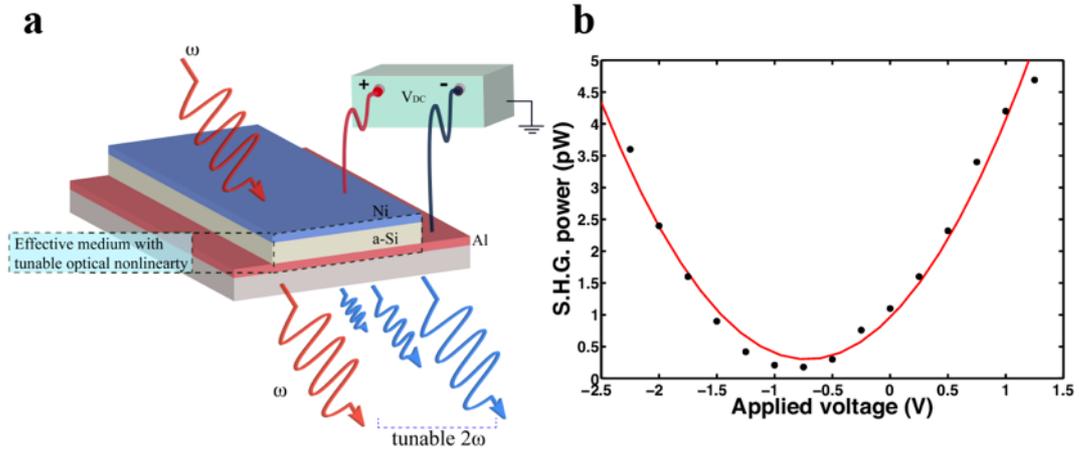

Fig. 4 (a) Schematic of an active MSM metamaterial. (b) Measured SHG intensity under variant DC voltage (dots), and the red fitting curve showing the quadratic dependency.

Table 1. Calculated components of the effective $\chi^{(2)}$ tensors.

| (pm/V) | *Ni / a-Si /Al* | *Pt / a-Si / Al* |
| --- | --- | --- |
| $\chi^{(2)}_{zzz}$ | 7.2 ± 1.5 | 8.9 ± 1.6 |
| $\chi^{(2)}_{xxz}$ | 1.9 ± 0.5 | 2.2 ± 0.8 |
| $\chi^{(2)}_{zxx}$ | 2.3 ± 0.8 | 2.3 ± 0.7 |